\documentclass[aps,prb,twocolumn,superscriptaddress]{revtex4}
\usepackage{graphics}
\usepackage{longtable}
\usepackage[epsfig]{graphicx}
\usepackage{epsfig}
\usepackage{amsmath}
\usepackage{float}
\usepackage{braket}
\usepackage{longtable}

\begin{document}
	
	\title{First principles investigation of quantum emission from hBN defects}
	
	\author{Sherif Abdulkader Tawfik$^{*}$}
	\affiliation{School of Mathematical and Physical Sciences, University of Technology Sydney, Ultimo, New South Wales 2007, Australia}
	
	\author{Sajid Ali$^{*}$}
	\affiliation{School of Mathematical and Physical Sciences, University of Technology Sydney, Ultimo, New South Wales 2007, Australia}
	\affiliation{Department of Physics, GC University Faisalabad,  Allama Iqbal Road, 38000 Faisalabad, Pakistan}
	
	\author{Marco Fronzi}
	\affiliation{School of Mathematical and Physical Sciences, University of Technology Sydney, Ultimo, New South Wales 2007, Australia}
	\affiliation{International Research Centre for Renewable Energy, State Key Laboratory of Multiphase Flow in Power Engineering, Xi'an Jiaotong University, Xi'an 710049, Shaanxi, China}
	
	\author{Mehran Kianinia}
	\affiliation{School of Mathematical and Physical Sciences, University of Technology Sydney, Ultimo, New South Wales 2007, Australia}
	
	\author{Toan Trong Tran}
	\affiliation{School of Mathematical and Physical Sciences, University of Technology Sydney, Ultimo, New South Wales 2007, Australia}
	
	\author{Catherine Stampfl}
	\affiliation{School of Physics, The University of Sydney, New South Wales, 2006, Australia}
	
	\author{Igor Aharonovich}
	\affiliation{School of Mathematical and Physical Sciences, University of Technology Sydney, Ultimo, New South Wales 2007, Australia}
	
	\author{Milos Toth}
	\affiliation{School of Mathematical and Physical Sciences, University of Technology Sydney, Ultimo, New South Wales 2007, Australia}
	
	\author{Michael J. Ford}
	\email{mike.ford@uts.edu.au}
	\affiliation{School of Mathematical and Physical Sciences, University of Technology Sydney, Ultimo, New South Wales 2007, Australia}

	\begin{abstract}
		Hexagonal boron nitride (hBN) has recently emerged as a fascinating platform for room-temperature quantum photonics due to the discovery of robust visible light single-photon emitters. In order to utilize these emitters, it is necessary to have a clear understanding of their atomic structure and the associated excitation processes that give rise to this single photon emission. Here we perform density-functional theory (DFT) and constrained DFT calculations for a range of hBN point defects in order to identify potential emission candidates. By applying a number of criteria on the electronic structure of the ground state and the atomic structure of the excited states of the considered defects, and then calculating the Huang-Rhys (HR) factor, we find that the C$_B$V$_N$ defect, in which a carbon atom substitutes a boron atom and the opposite nitrogen atom is removed, is a potential emission source with a HR factor of 1.66, in good agreement with the experimental HR factor. We calculate the photoluminescence (PL) line shape for this defect and find that it reproduces a number of key features in the the experimental PL lineshape.
	\end{abstract}		
	
	\maketitle

	\section{Introduction}
		
		Fluorescent defects in solids have attracted considerable attention as promising qubits for quantum information science.\cite{i1,i2,i3,i4} While the nitrogen vacancy ($N_V^{-}$) center in diamond has been widely investigated\cite{i5}, employing theoretical and experimental methodologies, defects in other wide band gap materials are far less understood.
		
		One particular example of defects of interest are color centers in hexagonal boron nitride (hBN). Defects in hBN have been shown to emit across the visible and the ultra violet spectral range. Moreover, many of them have been shown to exhibit non classical behaviour and act as single photon sources, important prerequisite for integrated quantum photonics.\cite{SP0,SP1,SP2,SP3_Fuchs,SP4_Wrachtrup,SP5_Basseet,highT,SP6,SP7,SP8,SP9,SP10,SP11,SP12,UV_defect} However, despite rigorous experimental studies, their origin remains under debate. Earlier suggestions included nitrogen and carbon related defects, oxygen defects, boron vacancies, and localized structural defects such as boundaries and dislocations.
		
		In this work we perform theoretical analysis of 35 different defects using density function theory (DFT). In particular, we focus on most probable candidates including oxygen, carbon, silicon, sulfur, flourine and phosphorus. We utilize the formalism of optically active centers in solids in order to evaluate the Huang-Rhys (HR) factor for the most likely defect structures, and compare our simulation with experimental results. Interestingly, we find that the carbon-antisite (C$_B$V$_N$, displayed in Fig. \ref{fig:fig_graphicdesign}(a)) defect matches well with the experimental PL data in terms of lineshape and its HR factor. The studied defects are shown in Figure 1. We chose the defects based on most likely impurities during the hBN growth and processing. For instance, it is known that carbon and oxygen are readily incorporated into the hBN during growth, while annealing of hBN for the emitter formation is done on a silicon substrate – that may cause silicon indiffusion into the hBN. 
	
		\section{Theoretical method}
		
		We performed density functional-theory (DFT) calculations of point defects in hBN in order to assign the observed single-photon emission to specific point defects. The calculations were performed using SIESTA\cite{SIESTA} and VASP.\cite{KressePRB1996} Our SIESTA and VASP calculations were performed using the generalized gradient approximation by Perdew, Burke and Ernzerhof (PBE).\cite{PBE} For the PBE calculations, we used a supercell that is comprised of $7\times 7$ unit cells of hBN (98 atoms), and we performed geometry relaxation for obtaining the ground state electronic structure with a $3\times 3\times 1$ \textbf{k}-grid in both DFT codes. SIESTA uses basis sets comprised of numerical atomic orbitals, and approximates the ionic potential in terms of Troullier-Martins\cite{TM} norm-conserving pseudopotentials. The auxiliary basis uses a real-space mesh with a kinetic energy cutoff of 750 Ry, and the basis functions are radially confined using an energy shift of 0.005 Ry. In the structural energy minimization, the internal coordinates are allowed to relax until all of the forces are less than 0.02 eV/\AA. A vacuum region of 20 \AA is added between the periodic images. In VASP, the valence electrons are separated from the core by use of projector-augmented wave pseudopotentials (PAW).\cite{BlochlPRB1994} The energy cut-off for the plane wave basis set is 500 eV, and the energy tolerance is $10^{-6}$ eV. In the structural energy minimization, the internal coordinates are allowed to relax until all of the forces are less than 0.01 eV/\AA. A vacuum region of 15\AA is added between the periodic images. The lattice constant of the hBN primitive cell is $a=2.515$ \AA~using both SIESTA and VASP, calculated on a $8\times 8\times 1$ \textbf{k}-grid. 
		
		After using ground-state DFT to explore the electronic structure of these defects, we calculate the PL line shapes using an extension of the quantum mechanical model, originally proposed by Huang and Rhys in their seminal paper,\cite{HR} and recently applied for GaN and ZnO\cite{4} and $NV^-$ diamond.\cite{njp}. This approach is has been applied to a wide range of luminescent defects (see Ref. \citenum{HendersonBook,5,10}). A key quantity in the Huang-Rhys method is the Huang-Rhys (HR) factor, which quantifies coupling between the electronic and vibronic states and gives a measure of the strength of the phonon side bands (PSB) relative to the zero-phonon line (ZPL). A technologically useful single photon source requires a small HR factor. Applying the HR theory requires the prediction of the excited state (ES) geometry and the phonon modes of the system including those associated with the presence of a defect. Excited state geometries can be calculated using constrained DFT, CDFT (or $\Delta$SCF \cite{DeltaSCF2}) where one electron is held in an excited state while the atomic positions are relaxed.
		
		We have performed the CDFT calculation using both SIESTA and VASP for comparison. CDFT for the defect structures while preserving the spin-polarization, assuming that the transition is spin-preserving. This is a widely applied technique for simulating optical excitation in molecules and crystals, and allows us to calculate the ZPL energy.\cite{DeltaSCF1,DeltaSCF2,DeltaSCF3,[PNAS]} In the SIESTA CDFT calculation, we have used a single $\Gamma$ point: Given that, in the ground state SIESTA calculation, the difference in total energy between using a single $k$ point and a $3\times 3$ k-point grid is $10^{-3}$ eV, this justifies using a single \textbf{k} point in excited state calculation and a $3\times 3\times 1$ \textbf{k}-grid in the ground state calculations. Phonon calculations were performed using the Phonopy code.\cite{phonopy}
		
		\begin{figure}[h]
			\includegraphics[width=90mm]{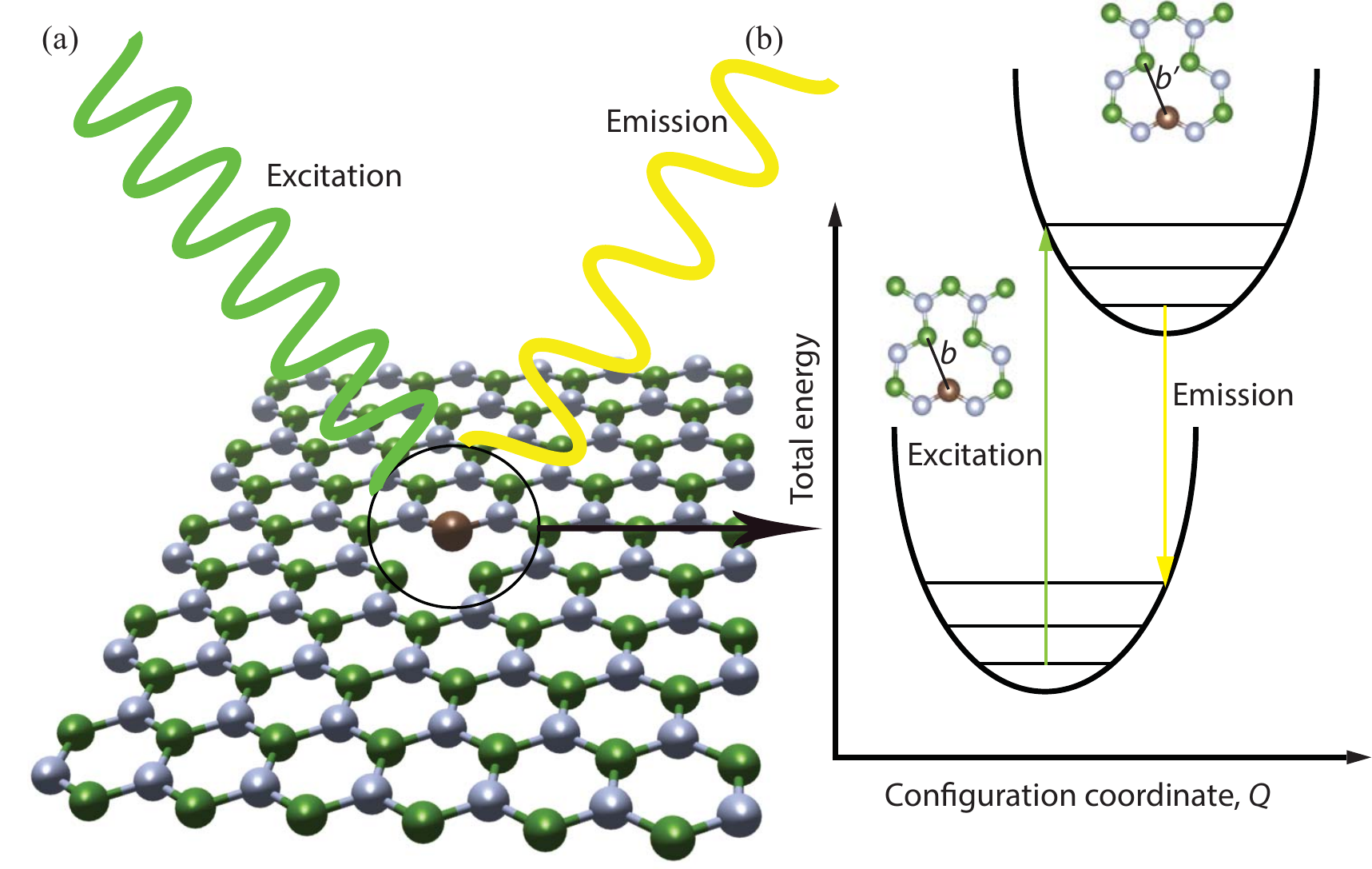}
			\caption{(Color online) (a) Schematic of the photoluminescence process, where the point defect is excited by green light to emit in the yellow range. The defect shown is the carbon-antisite defect, C$_B$V$_N$. (b) Schematic of the one-dimensional configuration coordinate diagram, displaying the total energy as a function of the configuration coordinate $Q$ for the excited state (top) and ground state (bottom) structures. The labels $b$ and $b^{'}$ denotes the C-B bond length in the ground and excited state structures, respectively.}
			\label{fig:fig_graphicdesign}
		\end{figure}
		
		The HR factor is calculated by applying the one-dimensional (1D) configuration coordinate formulation, where the atomic degrees of freedom are represented by a single coordinate, the configuration coordinate $Q$. Figure \ref{fig:fig_graphicdesign}(b) displays the 1D configuration diagram, showing the total energy as a function of $Q$ for the ground (lower curve) and excited (upper curve) states. The PL spectral function $L(\hbar\omega)$ is obtained following the generating function procedure,\cite{7,njp,4} and the electron-phonon spectral function, $S(\hbar\omega)$, the partial HR factor $S_k$, are calculated according to the procedure in Ref. \citenum{njp}, and are defined in the ESI$^{\dagger}$.
		
		In order to evaluate the Huang-Rhys factor from the experimental data, we utilize the relationship between $S$ and the intensity of the ZPL, $I_0$: 
		
		\begin{equation}
		S=-{\rm ln}(I^T_0/I^T)\quad,
		\label{eq:emperical_HR}
		\end{equation}
		
		\noindent where $I^T_0$ is the area under the ZPL line, while $I^T$ is the integral of the full PL lineshape.\cite{9} A measure of the extent of atomic structure change due to excitation is represented by the quantity $\Delta Q$ which is defined by the formula:
		
		\begin{equation}
		\label{DeltaQ}
		\Delta Q^2=\sum_{ai} m_a^{1/2}\left( R_{e,ai}-R_{g,ai}\right)\quad.
		\end{equation}
	
		\noindent where $a$ enumerates the atoms, $i= x,y,z$, $m_a$ is the atomic mass of species $a$, $R_{g/e,ai}$ is the position of atom $a$ in the ground/excited state, respectively.
		
		\section{Results and Discussion}
		
		\subsection{Defect Electronic Structures}

\begin{figure*}[h]
			\includegraphics[width=170mm]{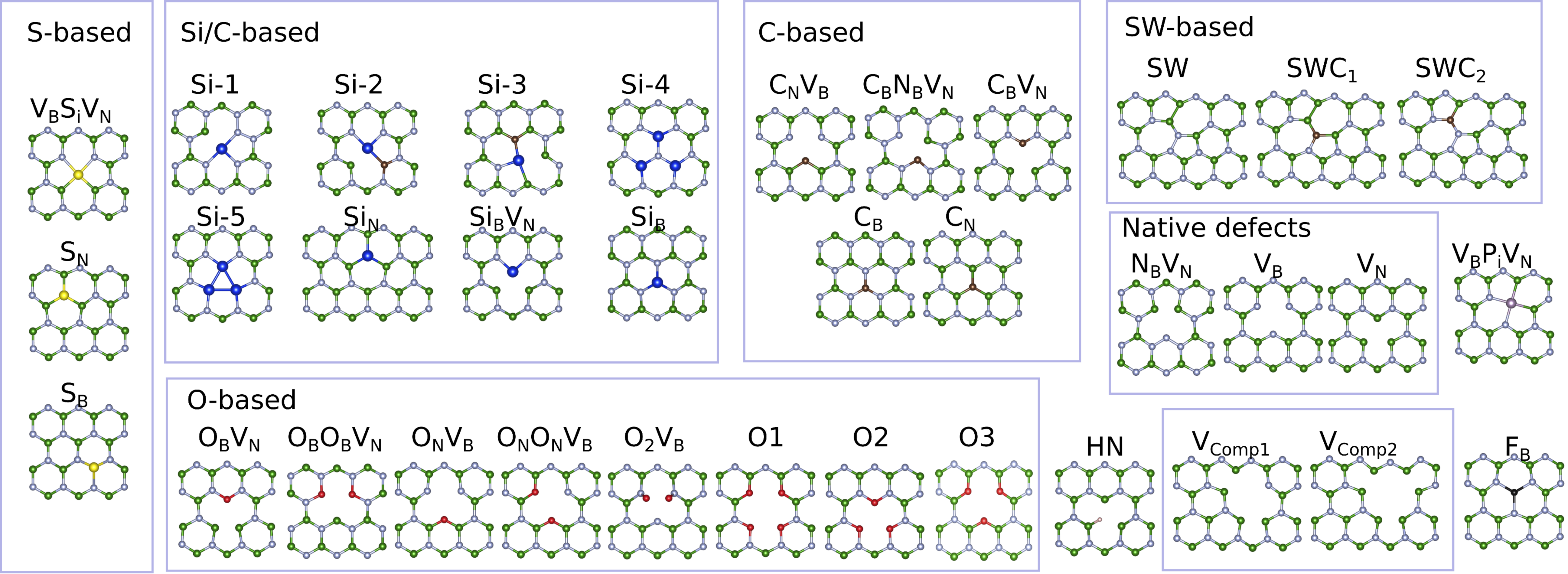}
			\caption{\label{fig:fig4} (Color online) Atomic structures of the defects in Tab. I. nitrogen atoms are represented by white spheres, boron green, oxygen red, silicon blue, carbon brown, sulfur yellow, fluorine black, phosphorus silver, hydrogen small white sphere. SWX are Stone-Wales defects, and V$_{CompX}$ are complex vacancy defects.}
		\end{figure*}

		The atomic structures of the 35 considered hBN defects are displayed in Fig. \ref{fig:fig4}. We consider several groups of defects: Si/C-defects, Stone-Wales defects (labeled SWC$_N$), C-based defects, O-based defects, native defects, S-based defects, complex defects, as well as a few other defects. We have applied the defect nomenclature where possible. For all of these defects, we have calculated the imaginary component of the dielectric constant and the band structure in the ground state (GS) in order to identify promising defects with emission properties consistent with the experiments. In Tab. 1, we display the results of the ground state calculations. The values of $E^{\uparrow\downarrow}_{T}$ correspond to characteristic peaks in the imaginary dielectric function (effectively peaks in the optical absorption spectrum) for the spin-up and spin-down channels respectively. To select potential emitters, we have applied 3 criteria, based on experimental observations, in order to narrow down the list of potential defects shown in Tab. 1:
		
		\begin{enumerate}	
			\item The stability of the emitters against annealing\cite{SP2} and high temperature\cite{highT} imply that the positions of the defect levels are within the band gap, and that none of them reside within, or close to, the bulk bands.\cite{[PNAS]}
			\item The observed emitters are polarized,\cite{SP1} and therefore we only focus on defects in which the defect excitation in the optical spectrum is also polarized.
			\item $E_{ZPL}$ is between 1.3 eV and 2.0 eV, which includes the range of ZPL energies reported in the literature.\cite{SP1,SP2,SP3_Fuchs,SP12} It is important to remember the limitations of DFT in this regard where it is well known that band-gaps in semiconductors are underestimated relative to experimental values.
		\end{enumerate}
		
\begin{figure}[h]
			\includegraphics[width=90mm]{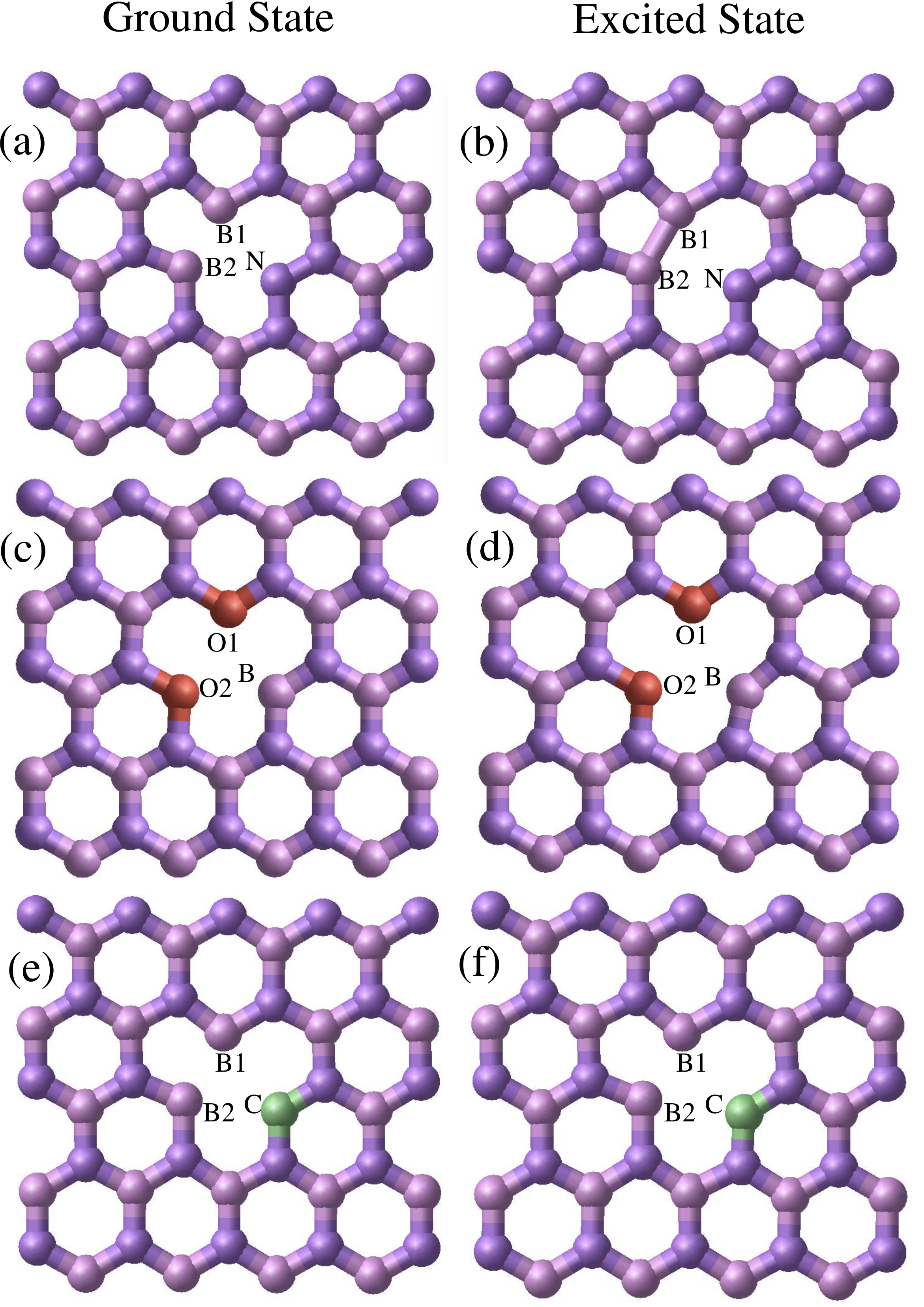}
			\caption{\label{fig:fig1} (Color online) The atomic structure of the ground state (GS) (a) and the excited state(ES) (b) of the N$_{B}$V$_{N}$ defect, GS (c) and ES (d) of the O$_B$O$_B$V$_{N}$ defect, and GS (e) and ES (f) of the C$_{B}$V$_{N}$ defect. Nitrogen atoms are displayed as violet spheres, B atoms as pink, C as green and O as red. The bond distances in (a) are N$-$B1/2: 2.59 \AA, B1$-$B2: 1.95 \AA, (b) N$-$B1/2: 2.64 \AA, B1$-$B2: 1.80 \AA, (c) B$-$O1/2: 2.52 \AA, O1$-$O2: 2.53 \AA, (d) B$-$O1/2: 2.68 \AA, O1$-$O2: 2.58 \AA, (e) C$-$B1/2: 2.53 \AA, B1$-$B2: 2.07 \AA and (f) C$-$B1/2: 2.62 \AA, B1$-$B2: 2.00 \AA.}
		\end{figure}

		A subset of the considered defects are presented in Tab. 2. In the defects of Tab. 2, $E^{\uparrow\downarrow}_{T}$ lies within the band gap, and criteria (1) and (2) are satisfied. Then, applying criterion (3) on the defects in Tab. 2, we found that there are three defect structures that could potentially emit single photons: the N-anti-site defect (N$_{B}$V$_{N}$) presented in Fig. \ref{fig:fig1}(a) ($\Delta Q=0.66$ amu$^{1/2}$ \AA), the O$_B$O$_B$V$_{N}$ defect ($\Delta Q=0.93$ amu$^{1/2}$ \AA) and the C$_B$V$_N$ defect ($\Delta Q=0.53$ amu$^{1/2}$ \AA). The calculations using VASP gave values for $E_{ZPL}$ that are very close to the values obtained using SIESTA: 2.01 eV, 1.85 eV and 1.33 eV for the N$_{B}$V$_{N}$, O$_B$O$_B$V$_{N}$ and C$_B$V$_N$ defects, respectively. The S$_B$, V$_N$C$_B$B$_N$, S$_V$B$_N$, Si-1 and Si-2 defects, although they satisfy the above four criteria, undergo large structural change as quantified by $\Delta Q$, as defined in Eq. \ref{DeltaQ} (for these defects, $\Delta Q=4.18$, 4.91, 1.45, 9.91 and 5.00 amu$^{1/2}$ \AA, respectively). Such large values for $\Delta Q$ would correspond to large HR values.

		\begin{table*}
			\label{tab:1}
			\centering
			\caption{The calculated electronic transition energy of the spin-up defect state, $E^{\uparrow}_{T}$, the spin-down defect state, $E^{\downarrow}_{T}$, whether the transition is from/to the bulk band, and whether the transition is optically allowed. All energies are in units of eV.}
			\begin{tabular}{|c|c|c|c|c|c|c|c|}
				\hline
				Defect	&	$E^{\uparrow}_{T}$	&Bulk bands?	& $E^{\downarrow}_{T}$	&Bulk bands? & Polarized	\\
				\hline
			C$_B$	&	-	&		&	-	&		&		\\
			C$_N$	&	-	&		&	-	&		&		\\
			C$_N$V$_B$	&	1.93	&		&	2.67	&		&	Yes	\\
			C$_B$V$_N$	&	1.44	&	No	&	-	&		&	Yes	\\
			C$_B$B$_N$V$_N$	&	1.00	& No &	1.00 & No & Yes \\							
			
			V$_B$PiV$_N$	&	3.41	&		&	-	&		&		\\
			
			F$_B$	&	2	&	No	&	2	&	No	&	No	\\
			
			S$_B$	&	2.26	&	No	&	-	&		&	Yes	\\
			V$_B$S$_i$V$_N$	&	2.72	&	No	&	2.72	&	No	&	Yes	\\
			S$_N$	&	-	&		&	-	&		&		\\
			
			N$_B$V$_N$	&	2.02	&	No	&	-	&		&	Yes	\\
			V$_B$	&	-	&		&	2.31	&	Yes	&		\\
			V$_N$	&	2.08	&	No	&	-	&		&	No	\\
			
			O$_N$V$_B$	&	-	&		&	-	&		&		\\
			O$_N$O$_N$V$_B$	&	-	&		&	-	&		&		\\
			O$_B$V$_N$	&	-	&		&	-	&		&		\\
			O$_B$O$_B$V$_N$	&	2.39	&	No	&	-	&		&	Yes	\\
			O$_2$V$_B$	&	2	&		&	0.87, 2.18	&		&	Yes	\\
			O-1	&	-	&		&	-	&		&		\\
			O-2	&	-	&		&	-	&		&		\\
			O-3	&	-	&		&	-	&		&		\\
			
			SW	&	3.55	&		&	3.55	&		&		\\
			SWC$_N$	&	2.94	&	No	&	-	&		&	Yes	\\
			SWC$_B$	&	2.02	&	Yes	&	-	&		&	No	\\
			
			V$_{Comp1}$	&	-	&		&	1.67	&	Yes	&		\\
			V$_{Comp2}$	&	-	&		&	-	&		&		\\

			Si$_B$	&	-	&		&	-	&		&		\\
			Si$_B$V$_N$	&	3.28	&	No	&	3.28	&	No	&	Yes	\\
			Si$_N$	&	-	&		&	-	&		&		\\
			Si-1	&	2.21	&		&	2.21	&		&		\\
			Si-2	&	2.2	&		&	-	&		&		\\
			Si-3	&	1.83	&	No	&	1.99	&	No	&		\\
			Si-4	&	0.42, 0.96	&	No	&	-	&		&		\\
			Si-5	&	3.44, 3.75	&		&	-	&		&		\\
			
			HN	&	1.75, 2.97	&	No	&	-	&		&		\\
				\hline
			\end{tabular}

		\end{table*}

		\begin{table}
			\centering
			\caption{The calculated ZPL energy $E_{ZPL}$ (in eV) and the $\Delta Q$ for the defects (defiend in Eq. \ref{DeltaQ}).}
			\begin{tabular}{|c|c|c|}
				\hline
				Defect	&	ZPL (spin-up)	&	$\Delta Q$	\\
				\hline
				SV$_{BN}$	&	1.94	&	1.45  \\
				S$_B$	&	1.26	&	4.18 \\
				Si$_B$V$_N$	&	0.94	& 4.00	 \\
				O$_B$O$_B$V$_N$	&	1.90	&	0.93 \\
				SWC$_N$	&	2.70	& 0.30	\\
				C$_N$V$_B$	&	1.11	&1.86	 \\
				C$_B$V$_N$	&	1.39	&	0.53 \\
				N$_B$V$_N$	&	2.12	&	0.66 \\
				Si-1	&	1.86	&	9.91 \\
				Si-2	&	2.03	&	5.00 \\
				Si-3	&	0.62	&5.02	 \\
				Si-5	&	3.81	&5.03	 \\
				V$_N$C$_B$B$_N$	&	2.16	&	4.91 \\
			
				\hline
			\end{tabular}
			\label{tab:2}
		\end{table}
		
		The atomic structure of the ground state (GS) and excited state (ES) structures of N$_{B}$V$_{N}$, O$_B$O$_B$V$_{N}$ and C$_{B}$V$_{N}$ defects are presented in Fig. 1(b), (d) and (f), respectively. In all of the structures, there is no out-of-plane displacement upon excitation, which means that the out-of-plane optical phonon modes do not contribute to the partial HR factor $S_k$. For N$_{B}$V$_{N}$, the main difference between the atomic structure of the obtained excited structure (cf. Fig. 1(b)) and that of the ground state structure (cf. Fig. 1(a)) is that the B-B bond distance drops by 0.14 \AA; for O$_B$O$_B$V$_{N}$, the B$-$O1/2 stretches by 0.16 \AA; for C$_{B}$V$_{N}$, which undergoes the least atomic displacement, C$-$B1/2 stretches by 0.09 \AA. The atomic structure of the GS for all of the considered structures is available in the ESI$^{\dagger}$. 
		
		\subsection{Photoluminescence lineshapes and the Huang-Rhys factors}
		
		In our search for potential emitter structures, we have calculated the theoretical HR factors for the four emitters considered in the previous section using SIESTA. The values are: 4.49 for N$_{B}$V$_{N}$, 6.74 for O$_B$O$_B$V$_{N}$ and 1.66 for C$_{B}$V$_{N}$. That is, the emitter whose theoretical HR factor agrees with the experimental HR factor value of the emitter (which we calculated as $S$=1.66 using Eq. \ref{eq:emperical_HR}) is the C$_{B}$V$_{N}$. The HR factors of the other two emitters are too high compared with the available experimental HR value.
		
\begin{figure}[h]
			\includegraphics[width=90mm]{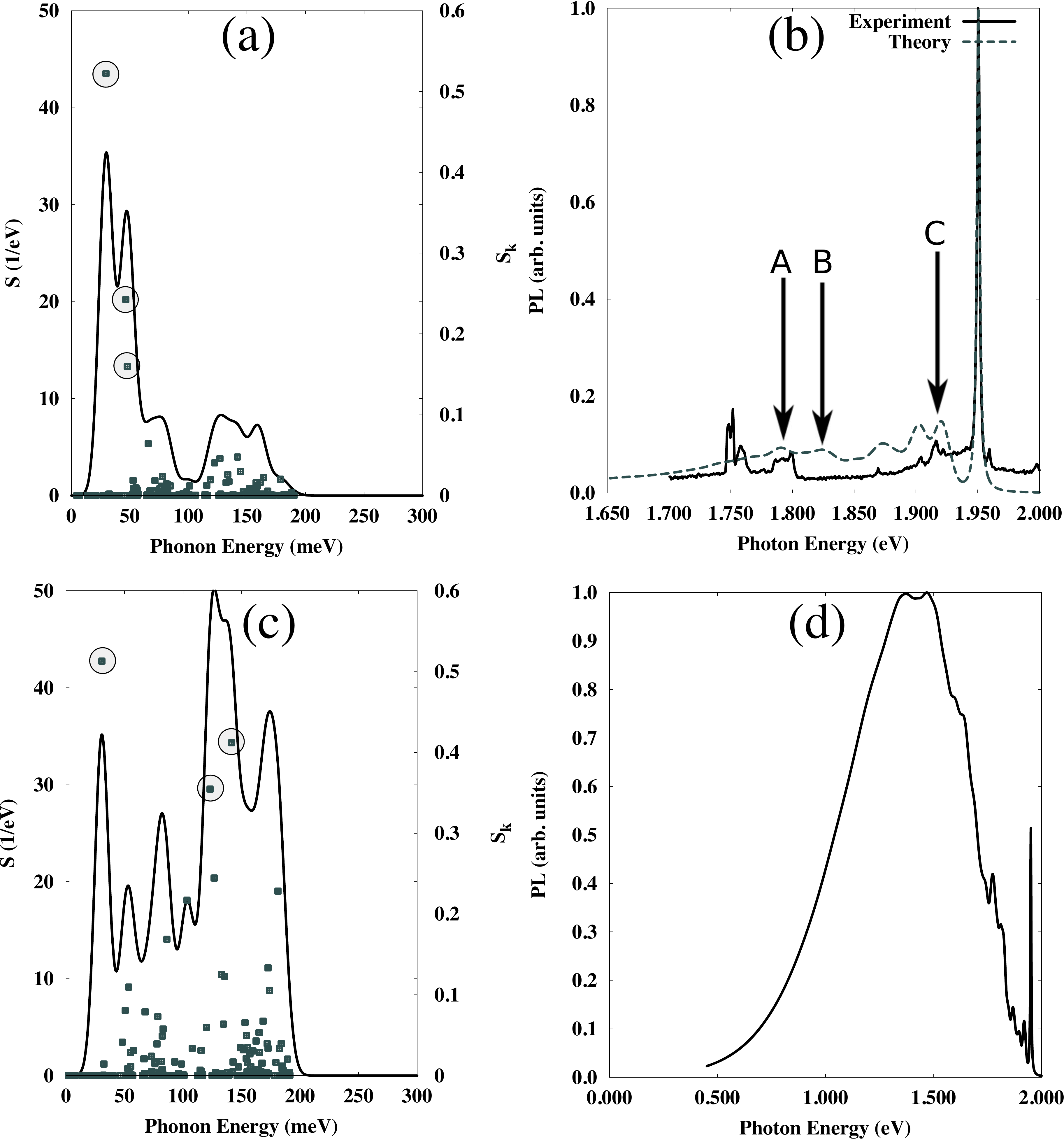}
			\caption{\label{fig:PL_Lineshape_Comparison} (Color online) For the C$_{B}$V$_{N}$ defect, (a) the set of partial HR factors, $S_k$ (displayed as sqare-shaped points), the approximate partial HR function, $S(\hbar\omega)$ (displayed as solid curves), and (b) the experimental and theoretical photoluminescence lineshapes. The same are displayed for N$_{B}$V$_{N}$ in (c) and (d), except that (d) does not display an experimental lineshape. The phonon modes labelled with circles in (a) and (c), which have the most significant contribution to the HR value, are displayed in Fig. \ref{fig:fig5}.}
		\end{figure}

		
		The partial HR factor $S_k$, the spectral function of electron-phonon coupling $S(\hbar\omega)$ and the PL lineshape $L(\hbar\omega)$ are displayed in Fig. \ref{fig:PL_Lineshape_Comparison} for the two defects, C$_{B}$V$_{N}$ and N$_{B}$V$_{N}$. The theoretical lineshape of C$_{B}$V$_{N}$ is superimposed on the experimental lineshape with a ZPL at 1.951 eV, which has been measured at 4K (Supplementary Information). The two spectral functions terminate at 200 meV, which is the maximum of the hBN phonon energy. The value of $\gamma$ has been adjusted in order to fit the experimental spectrum in Fig. \ref{fig:PL_Lineshape_Comparison}(b), and the same value is used in Fig. \ref{fig:PL_Lineshape_Comparison}(d). In the experimental spectrum of Fig. \ref{fig:PL_Lineshape_Comparison}(a), there are PSB peaks at $1.92$ eV, $1.8$ eV, $1.76$ eV, and $1.75$ eV. Using Eq. \ref{eq:emperical_HR}, the HR factor is 1.66. This spectrum resembles that of the 4.1 eV color center of hBN reported in Ref. \citenum{UV_defect} (measurement made at 10K): a broad low-energy feature right next to the ZPL peak, a peak at the maximum phonon energy of 200 meV and fine structure in between. Although the resemblence between the lineshapes does not imply that the emitting sources have any relationship, it helps to identify he sharp peak at $1.75$ eV (200 meV away from the ZPL). This peak has been identified by the authors of Ref. \citenum{UV_defect} as a zone-center longitudinal optical phonon. Such phonons scale as $1/k$, where $k$ is the phonon wave vector, and therefore the electron-phonon matrix element of these phonons diverge, leading to such intense peak.\cite{UV_defect}
		
		We can observe that the spectrum of the C$_{B}$V$_{N}$ defect has two high-energy features, A and B, corresponding to the two high energy features in the experimental spectrum, but blue-shifted by $\sim 20$ meV from the corresponding positions in the experimental spectrum. In addition, the theoretical spectrum in Fig. \ref{fig:PL_Lineshape_Comparison}(b) exhibits a low energy peak, C, that is $\sim 10$ meV away from the 1.90 eV peak in the experimental spectrum. On the contrary, the N$_{B}$V$_{N}$ lineshape considerably deviates from the experimental lineshape: there is a large peak that is $\sim$ 470 meV away from the ZPL. This information strongly suggests that the C$_{B}$V$_{N}$ defect is one of the SPE sources in hBN.
		
\begin{figure}[h]
			\includegraphics[width=90mm]{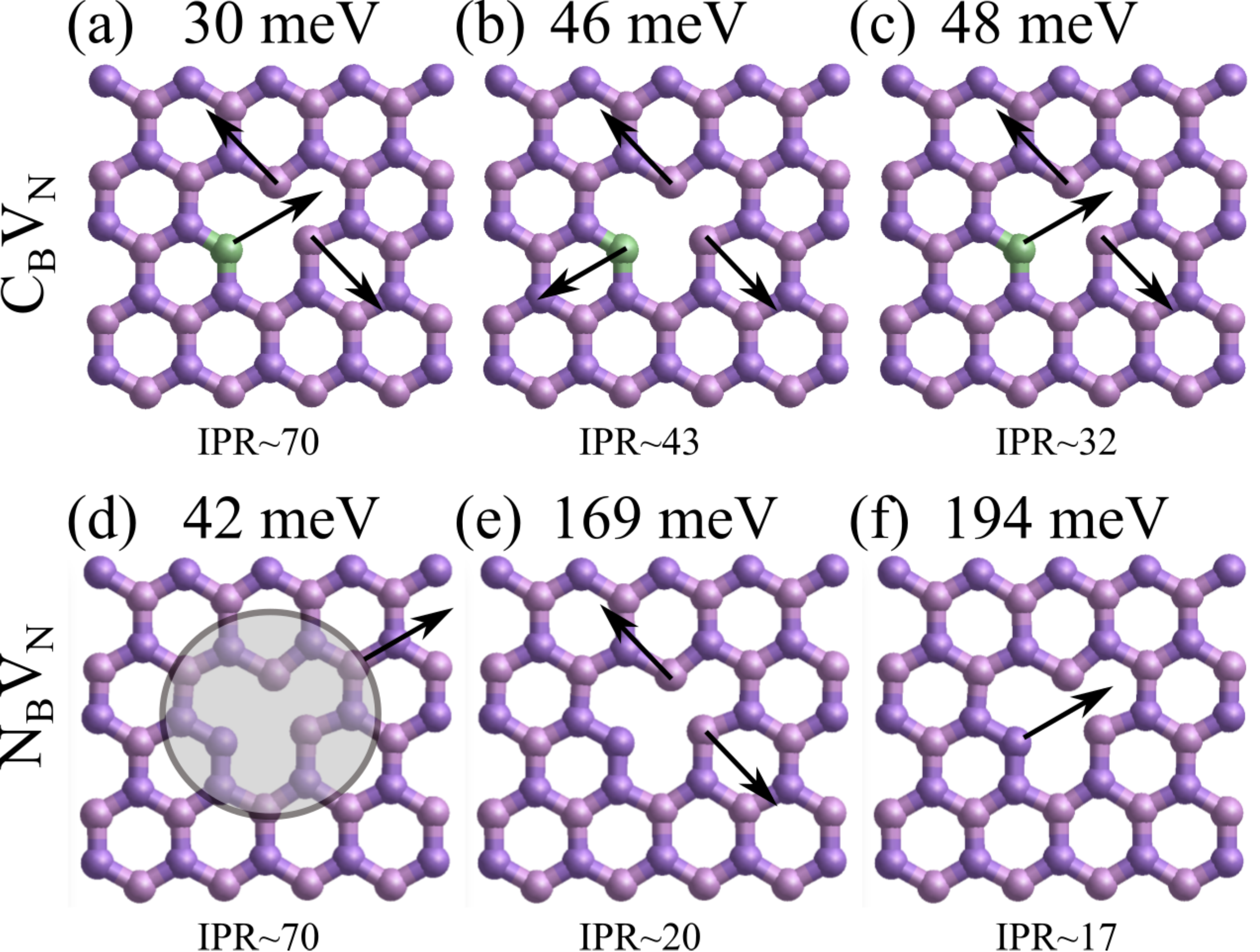}
			\caption{\label{fig:fig5} (Color online) A schematic diagram for the force vectors of the three phonon modes which have the largest contribution to the HR factor for the (a-c) C$_{B}$V$_{N}$ defect and the (d-f) N$_{B}$V$_{N}$ defect. In (d), the atoms enclosed by the circule move collectively.}
		\end{figure}

		Next, we analyse the localization-delocalization of phonon modes that contribute to the PL lineshape of the two defects. Although the concept of degree of localization is not well defined for complex structures, it is possible to roughly quantify the degree of localization based on the inverse participation ratio, or IPR\cite{njp}. This quantity gives the number of atoms that vibrate in a given normal mode, and ranges from 1 (a single atom vibrating) up to the total number of atoms in the supercell (97 in the case of the three defect structures). For the case of the C$_{B}$V$_{N}$ defect, we analyze the $S_k$ function displayed in Fig. \ref{fig:PL_Lineshape_Comparison}(a). Three modes represent 55.6\% of the total HR factor. The largest contribution is from a mode at 30 meV with $IPR_k=70$, and represents 31.5\% of the total HR factor. Then, a slightly localized mode at 46 meV with $IPR_k=43$ and represents 14.6\%. Then a slightly localized mode at 48 meV with $IPR_k=32$ and represents 14.6\%. These three phonon modes are illustrated in Fig. \ref{fig:fig5}(a,b,c). Both the 30 meV and the 48 meV phonon modes are defect breathing modes, in which the atoms surrounding the defect oscillate along the dipole direction of the defect. Therefore, they are infrared active modes, because the dipole moment of the defect changes. The 46 meV phonon mode is also a defect breathing mode, but the defect atoms in this mode oscillate radially around the defect site; that is, it is a symmetric breathing mode which is Raman active. A few localized modes with $IPR_k<15$ exist at low phonon energy and with negligible contribution to the total HR factor. The localized mode with highest contribution is at 142 meV, with $IPR_k=12$ and represents 2.9\% of the total HR factor. The high energy peaks that are responsible for the $S(\omega)$ peak at 150 meV in \ref{fig:PL_Lineshape_Comparison}(a) collectively represent 6.8\% of the total HR factor and with $32<IPR_k<58$. This information indicates that phonon modes that contribute to photoluminescence in the C$_{B}$V$_{N}$ defect are not local modes. This is in agreement with the conclusion in Ref. \citenum{UV_defect} for the case of UV luminescence.
		
		For the case of the N$_{B}$V$_{N}$ defect, the peak at 42 meV is a delocalized phonon mode in which the defect site move along the dipole direction, accompanied by the mobility of the majority of the atoms surrounding the defect. This mode has the largest partial HR factor (11.4\% of the total HR factor) with an IPR value of $\sim$ 70. The two modes with next-largest HR contribution have energy values 194 and 169 meV and contributes 17\% of the total HR factor, and have IPR values of $\sim$ 17 and $\sim$ 20, where the largest displacements originate from the atoms surrounding the defect site. These three phonon modes are illustrated in Fig. \ref{fig:fig5}(d,e,f). In the 194 meV mode, the N atom in the defect sites streches further into the defect site (towards the two B atoms), while in the 169 meV mode the two B atoms in the defect site strech perpendicular to the dipole direction(where each B atom strech in a direction opposite to the other). It is the strong electron-phonon coupling in these two high energy defect phonon modes that is responsible for the large PSB in the lineshape (cf. Fig. \ref{fig:PL_Lineshape_Comparison}(d)), and hence the large HR factor.

		Having applied the quantum theory of F-centers to potential hBN single-photon emitters, it is important to highlight that this theory relies on a number of assumptions and has emerged from the analysis of three-dimensional defect geometries. Therefore, it requires some developments to expand its usefulness for two-dimensional van der Waals crystals, whose physical properties are different from those of three dimensional crystals. If we accept the expression for $S_k$ at face value, then the coupling between electronic excitation and out-of-plane optical phonons will fail to be achieved by a defect in which the optical excitation induces a movement of the atoms in the $z$-direction. Algebraically speaking, this problem stems from the very definition of $S_k$ in terms of the product $R_{ia,g/e} \Delta r_{ia,k}$, because it depends on the actual coordinate value of $R_{ia,g/e}$. That is, the atoms whose $z$-axis coordinate is zero (such as graphene, hBN, etc) will have $R_{ia,g} \Delta r_{ia,k}=0$. If the atoms are displaced along the $z$-axis due to optical excitation, then the present theory will incorporate the contribution of optical out-of-plane phonon modes in the electron-phonon coupling by virtue of $R_{ia,g} \Delta r_{ia,k}\neq 0$. However, as long as the atoms do not move along the $z$-axis, all of the out-of-plane optical modes will not contribute to the electron-phonon coupling, and hence, to the luminescence process. To determine whether this is physically valid, more research on extending the classical F-center theory to 2D materials must be performed. This is cuurently in progress.
		
		\section{Conclusion}
		We applied density-functional theory and constrained-density functional theory to hBN in order to predict the luminescence properties of point defects in this material. The ground state electronic and geometric properties were deterimined for 35 different defect structures, from which a list of 13 defects were found to satisfy a number of criteria. Only one defect out of this list, the C$_{B}$V$_{N}$ defect, was found to reproduce the experimental HR factor, and its theoretical PL lineshape is in reasonable agreement with the experimental lineshape. The theoretical analysis of the phonon modes contributing to the PL lineshape of both the C$_{B}$V$_{N}$ and N$_{B}$V$_{N}$ defect shows that delocalized modes dominate the PL lineshape in C$_{B}$V$_{N}$, which is not the case in the N$_{B}$V$_{N}$ defect in which the phonon modes with highest contribution are strongly localized. Such localization results in a large phonon-side band contribution in the N$_{B}$V$_{N}$ defect. Our results provide a fundamental insight in the evolving field of two-dimensional quantum emitters, and will be useful in the engineering and control of such emitters.
		
		\section{Acknowledgements}
		
		S. A. T. and S. A. have contributed equally to this work. This research was funded by the Australian Government through the Australian Research Council (​ARC DP16010130). The theoretical calculations in this research were undertaken with the assistance of resources from the National Computational Infrastructure (NCI), which is supported by the Australian Government. The theoretical calculations in this work were also supported by resources provided by the Pawsey Supercomputing Centre with funding from the Australian Government and the Government of Western Australia. For the experimental measurements, financial support from the Australian Research Council (via DP140102721, DE130100592), and the Asian Office of Aerospace Research and Development grant FA2386-15-1-4044 is gratefully acknowledged.

\end{document}